\documentclass[graphicx,showpacs,aps,twocolumn]{revtex4}
\usepackage{graphicx}

\begin{document}

\title{ Faithful entanglement sharing for quantum communication against collective noise\footnote{Published in Int J Theor Phys \textbf{51}, 2346-2352 (2012)}}
\author{Hui-Chong Niu$^1$, Bao-Cang Ren$^1$, Tie-Jun Wang$^{1,2}$, Ming Hua$^1$,
and Fu-Guo Deng$^1$\footnote{Corresponding author. Email address: fgdeng@bnu.edu.cn.}}
\address{$^1$ Department of Physics, Applied Optics Beijing Area Major
Laboratory, Beijing Normal University, Beijing 100875, China\\
$^2$
Department of Physics, Tsinghua University, Beijing 100084, China}

\date{\today }

\begin{abstract}
We present an economical setup for faithful entanglement sharing
against collective noise. It is composed of polarizing beam
splitters, half wave plates, polarization independent wavelength
division multiplexers, and frequency shifters. An arbitrary qubit
error on the polarization state of each photon in a multi-photon
system caused by the noisy channel can be rejected, without
resorting to additional qubits, fast polarization modulators, and
nondestructive quantum nondemolition detectors. Its success
probability is in principle 100\%, which is independent of the noise
parameters, and it can be applied directly in any one-way quantum
communication protocol based on entanglement.
\end{abstract}
\pacs{03.67.Pp, 03.67.Hk} \maketitle

\section{Introduction}

Quantum entanglement is an important resource in quantum information
processing and transmission, such as quantum computation \cite{qec}
and quantum key distribution (QKD) \cite{RMP}. In optical QKD
protocols, the two remote parties, say the sender Alice and the
receiver Bob, exploit entangled photon pairs to set up their quantum
channel for creating a sequence of random bits as their private key
\cite{Ekert91,BBM92,Long-Liu,dengpra}, in particular in a
long-distance quantum communication. However, entangled photon
systems will inevitably interact with their environments, which will
make them decoherence. For instance, the thermal fluctuation,
vibration, and the imperfection of the fiber will affect the
polarization of photons transmitted in an optical fiber. For
accomplishing the task of quantum communication, people have
proposed some interesting methods to overcome the effect of the
channel noise, such as phase coding \cite{phase}, quantum error
correct code (QEC) \cite{qec}, entanglement purification \cite{ep1},
error rejection \cite{pla343,apl91,prl95,lixhjpb,dengqic},
decoherence-free subspace (DFS) \cite{dfs1,dfs2,dfs4}, and so on. In
QKD protocols with phase coding \cite{phase}, the fluctuation of
optical fiber is passively compensated with  the Faraday
orthoconjugation effect via a two-way quantum communication.
However, the use of two-way quantum communication makes the system
vulnerable to the Trojan horse attack. Entanglement purification
will consume exponentially the quantum resource for obtaining a
subset of maximally entangled photon pairs from a set of less
entangled systems \cite{ep1} or it will resort to hyperentanglement
which is the entanglement in several degrees of freedom of photons
\cite{ep6}, such as polarization, frequency, spatial mode, and so
on. Although QEC and DFS methods can be used to suppress the effect
of channel noise effectively, they are sensitive to channel loss and
need much resource \cite{shengpra}.

Usually, the fluctuation in an optical fiber is slow in time, so
that the alteration of the polarization is considered to be the same
over the sequence of photons \cite{prl95}. This feature provides
people a good way to design quantum error-rejection protocols. For
instance, Yamamoto \emph{et al.} \cite{prl95} proposed an
error-rejecting scheme for the faithful transmission of a
single-photon polarization state over a collective-noise channel,
resorting to a reference single photon in a fixed state in 2005. Its
success probability is in principle 1/16 without two-qubit
operations \cite{xihanOC}. Subsequently, Kalamidas \cite{pla343}
proposed two schemes to reject and correct arbitrary qubit errors
without additional particles, but fast polarization modulators. In
2007, Li \emph{et al.} \cite{apl91} also proposed a faithful qubit
transmission scheme against a collective noise without ancillary
qubits. Its success probability is 50\% with only linear optical
elements in a passive way.  These error-rejection protocols can also
be used for the distribution of entangled photon pairs between two
remote parities. However, their success probability is very low. In
2010, Sheng and Deng \cite{shengpra} presented an interesting scheme
for quantum entanglement distribution  over an arbitrary
collective-noise channel.  In 2011, Li and Duan \cite{lixhjpb}
proposed an interesting scheme for efficient polarization qubit
transmission against a collective polarization noise, resorting to
the frequency degree of freedom of a single photon.

In fact, the polarization of photons is, on the one hand, sensitive
to channel noise. The frequency of photons suffers less from channel
noise as its alteration requires a nonlinear interaction between
photon and an optical fiber, which takes place with a negligible
probability. The previous experiments showed that the polarization
entanglement is quite unsuitable for transmission over distances of
more than a few kilometers in an optical fiber \cite{RMP}. For
example, Naik \emph{et al.} \cite{naik} observed the quantum bit
error rate (QBER) increase to 33\% in the experimental
implementation of the six-state protocol \cite{sixstate1,sixstate2}
over only a few meters. For frequency coding, for example, the
Besancon group performed a key distribution over a 20-km single-mode
optical-fiber spool and they recorded an
 QBER contribution of approximately 4\%. They estimated that
2\% could be attributed to the transmission of the central frequency
by the Fabry-Perot cavity \cite{frequency}. On the other hand, the
measurement on the polarization states of photons is easier than
that on the frequency states, which is important in quantum
communication protocols in which the parties resort to at least two
bases for checking eavesdropping, such as the
Bennett-Brassard-Mermin 1992 (BBM92) QKD protocol \cite{BBM92} and
the Hillery-Bu\v{z}k-Berthiaume (HBB) quantum secret sharing (QSS)
protocol \cite{QSS}. The alternation of basis of polarization equals
a Hadamard operation and it can be completed with a $\lambda$/4
plate, but it is difficult for frequency.

In this paper, we will present a new setup for faithful entanglement
sharing against collective noise, without resorting to additional
qubits, fast polarization modulators, and nondestructive quantum
nondemolition detectors based on nonlinear media. This setup is
composed of polarizing beam splitters,  half wave plates,
polarization independent wavelength division multiplexers, and
frequency shifters, which is feasible with current technology. An
arbitrary qubit error on the polarization state of each qubit caused
by the noisy channel can be rejected with the success probability
100\% in principle, which is independent of the noise parameters.
This setup has good applications in any one-way quantum
communication protocol based on entanglement for rejecting the
errors caused by the collective noise in channel. We will discuss
its applications in the famous BBM92 QKD scheme and the HBB QSS
protocol.

\begin{figure}[!h]
\begin{center}
\includegraphics[width=8cm,angle=0]{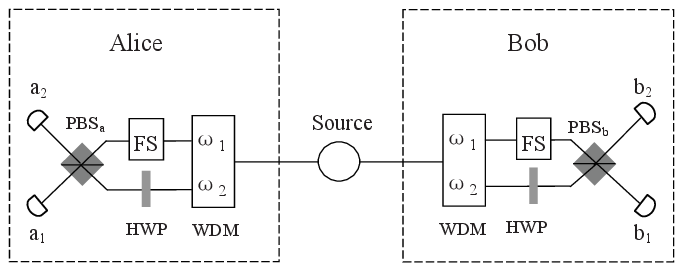}
\caption{Our setup  for faithful entanglement sharing over a
collective-noise channel.  Two photons $a$ and $b$ are transmitted
to Alice and Bob, respectively. WDM represents a polarization
independent wavelength division multiplexer which will lead the
photons to different spatial modes according to their frequencies.
FS is a frequency shifter which is used to complete the frequency
shift from $\omega_1$ to $\omega_2$. HWP represents a half-wave
plate which is used to accomplish the transformation $\vert H\rangle
\leftrightarrow \vert V\rangle$. PBS is a polarizing beam splitter.}
\label{fig1}
\end{center}
\end{figure}


\section{Faithful entanglement sharing over a collective noise}

The setup for implementation of the present faithful entanglement
sharing between two parties is shown in Fig.\ref{fig1}. Suppose that
there is an entangled quantum  source placed at a point between the
two parties in quantum communication, say Alice and Bob. The source
emits an entangled photon pair $ab$ in the state $|\Psi\rangle_{ab}$
each time. Here
\begin{eqnarray}
|\Psi\rangle_{ab}&=&\frac{1}{\sqrt{2}}|H\rangle_{a}|H\rangle_{b}
(|\omega_{1}\rangle|\omega_{2}\rangle
+|\omega_{2}\rangle|\omega_{1}\rangle)_{ab} \nonumber\\
&\equiv&\frac{1}{\sqrt{2}}(\vert H \rangle_{\omega_{1}} \vert H
\rangle_{\omega_{2}} + \vert H \rangle_{\omega_{2}} \vert H
\rangle_{\omega_{1}})_{ab},
\end{eqnarray}
where $|H\rangle$ and $|V\rangle$ present the horizontal
polarization and the vertical polarization of photons, respectively.
$|\omega_{1}\rangle$ and $|\omega_{2}\rangle$ are two different
frequency modes of  photons. The subscripts $a$ and $b$ mean that
the two photons transmitted are distributed to Alice and Bob,
respectively. There ate two channels for the two parties. One is
connecting the quantum entangled source and Alice, and the other is
used to link the source with Bob. Suppose that the collective noise
in these two channels have the same form but different noise
parameters, and the alternation of the polarization caused by the
noisy channels can be described as:
\begin{eqnarray}
\vert H \rangle_a  &^{\underline{\; noise_a }}\rightarrow& \alpha
\vert H \rangle +\beta \vert V \rangle,\nonumber\\ \vert H \rangle_b
& ^{\underline{\; noise_b \;}}\rightarrow&   \delta \vert H \rangle
+\gamma\vert V \rangle,
\end{eqnarray}
where
\begin{eqnarray}
\vert \alpha \vert ^{2}+ \vert \beta \vert^{2}=1, \;\;\;\;\;\; \vert
\delta \vert^{2} + \vert \gamma \vert^{2}=1.
\end{eqnarray}

After being transmitted through optical-fiber channels, each photon
of the entangled pair is influenced by the collective noise in its
channel. That is, the evolution of the entangled state through the
noisy channels can be written as:
\begin{eqnarray}
\vert \Psi \rangle_{ab} &^{\underline{  noise }}\rightarrow&
\frac{1}{\sqrt{2}}\{\left( \alpha\vert H \rangle_{\omega_{1}} +
\beta \vert V \rangle_{\omega_{1}}\right)\left(\delta \vert H
\rangle_{\omega_{2}}
+ \gamma \vert V\rangle_{\omega_{2}}\right) \nonumber\\
&& \;\; + \;\; \left(\alpha \vert H \rangle_{\omega_{2}} +\beta
\vert V \rangle_{\omega_{2}}\right)\left(\delta \vert H
\rangle_{\omega_{1}}
+\gamma \vert V\rangle_{\omega_{1}}\right)\}\nonumber\\
&=&\frac{1}{\sqrt{2}}\{\alpha\delta\left(\vert H
\rangle_{\omega_{1}} \vert H \rangle_{\omega_{2}} +\vert H
\rangle_{\omega_{2}} \vert H \rangle_{\omega_{1}}\right) \nonumber\\
&& \;\; + \;\; \alpha\gamma\left(\vert H \rangle_{\omega_{1}} \vert
V \rangle_{\omega_{2}} +\vert H
\rangle_{\omega_{2}} \vert V \rangle_{\omega_{1}}\right)\nonumber\\
&&\;\; + \;\; \beta\delta\left(\vert V \rangle_{\omega_{1}} \vert H
\rangle_{\omega_{2}} + \vert V \rangle_{\omega_{2}}\vert H
\rangle_{\omega_{1}}\right) \nonumber\\
&& \;\; + \;\;  \beta\gamma\left(\vert V \rangle_{\omega_{1}} \vert
V \rangle_{\omega_{2}} +\vert V \rangle_{\omega_{2}} \vert V
\rangle_{\omega_{1}}\right)\}.
\end{eqnarray}

After the noisy channel, each of the photons will pass through a
polarization independent wavelength division multiplexer (WDM) which
guides photons to different spatial modes, according to their
frequencies. That is, the photon with the frequency $\omega_1$ is
led to the upper spatial mode  and the photon with the frequency
$\omega_2$ is led to the lower spatial mode, respectively. This
device can also be substituted by a frequency beam splitter
\cite{r9} or a fiber bragg grating \cite{r10,r11}. After the WDMs,
Alice and Bob adjust the photons transmitted through the upper
spatial modes  from the frequency state $\omega_1$ to $\omega_2$, so
as to make the quantum states in the two outports of each WDM have
the same frequency. Also, they transform the polarization states of
the photons transmitted through the lower spatial modes. Obviously,
the state of the system after the two PBSs (i.e., PBS$_a$ and
PBS$_b$) becomes
\begin{eqnarray}
&^{\underline{ PBSs}}\rightarrow&
\frac{1}{\sqrt{2}}\{\alpha\delta\left(\vert H \rangle \vert V
\rangle + \vert V \rangle  \vert H \rangle \right)_{a_1b_1} \nonumber\\
&&\;\; + \;\;\; \alpha\gamma\left(\vert H \rangle  \vert H \rangle +
\vert V
\rangle  \vert V \rangle \right)_{a_1b_2}\nonumber\\
&&\;\; + \;\;\; \beta\delta\left(\vert V \rangle  \vert V \rangle +
\vert H \rangle \vert H \rangle \right)_{a_2b_1}  \nonumber\\
&&\;\; + \;\;\; \beta\gamma\left(\vert V \rangle  \vert H \rangle  +
\vert H \rangle \vert V \rangle \right)_{a_2b_2}\}.
\end{eqnarray}
Here $a_{1}$ and $b_{2}$ are two output ports of  PBS$_a$, and
$b_{1}$ and $b_{2}$ are those of PBS$_b$. In essence, these two PBSs
filter the polarization states of photons with different spatial
modes. That is, Alice and Bob can determine the polarization states
of their photons according to their spatial modes by postselection.
There are four combination modes of the output ports for the two
entangled photons $ab$. That is, they emit from the output ports
$a_1b_1$, $a_1b_2$, $a_2b_1$, or $a_2b_2$ and they are in the
polarization-entangled states $\vert
\psi^+\rangle_{a_1b_1}=\frac{1}{\sqrt{2}}\left(\vert H \rangle \vert
V \rangle + \vert V \rangle  \vert H \rangle \right)_{a_1b_1}$,
$\vert \phi^+\rangle_{a_1b_2}=\frac{1}{\sqrt{2}} \left(\vert H
\rangle  \vert H \rangle + \vert V \rangle  \vert V \rangle
\right)_{a_1b_2}$, $\vert
\phi^+\rangle_{a_2b_1}=\frac{1}{\sqrt{2}}\left(\vert V \rangle \vert
V \rangle + \vert H \rangle \vert H \rangle \right)_{a_2b_1}$, and
$\vert \psi^+\rangle_{a_2b_2}=\frac{1}{\sqrt{2}}\left(\vert V
\rangle \vert H \rangle  + \vert H \rangle  \vert V \rangle
\right)_{a_2b_2}$, respectively. The total success probability for
this entanglement distribution scheme is $|\alpha\delta|^2 +
|\alpha\gamma|^2 + |\beta\delta|^2 + |\beta\gamma|^2=1$ in
principle, independent of the noise parameters.

By far, we have discussed our setup for entanglement sharing over a
collective-noise channel in which the noise affects the polarization
states of the photons but not entangles with them. In detail, the
polarization states of the photons transmitted are some pure states.
In general, the polarization state of an entangled photon system
after the transmission over a noisy channel becomes a mixed one.
Suppose a two-photon system after being transmitted is in a mixed
state as follows
\begin{eqnarray}
\vert \Psi \rangle_{ab} &^{\underline{  noise }}\rightarrow&
\rho_{ab}=\rho_p \cdot \rho_f,
\end{eqnarray}
where
\begin{eqnarray}
\rho_p &=& F_1 \vert HH\rangle\langle HH\vert + F_2 \vert
HV\rangle\langle HV\vert\nonumber\\
& +& F_3 \vert VH\rangle\langle VH\vert + F_4 \vert VV\rangle\langle
VV\vert
\end{eqnarray}
 and
\begin{eqnarray}
\rho_f =\frac{1}{2} (|\omega_{1}\rangle|\omega_{2}\rangle
+|\omega_{2}\rangle|\omega_{1}\rangle)_{ab}(\langle\omega_{1}|\langle\omega_{2}|
+\langle\omega_{2}|\langle\omega_{1}|).
\end{eqnarray}
This state can be viewed as a probabilistic mixture of four pure
states: with a probability of $F_1$, $F_2$, $F_3$, or $F_4$,  the
photon pair in the state $\frac{1}{\sqrt{2}}\vert H\rangle_a \vert
H\rangle_b (|\omega_{1}\rangle|\omega_{2}\rangle
+|\omega_{2}\rangle|\omega_{1}\rangle)_{ab}$,
 $\frac{1}{\sqrt{2}}\vert H\rangle_a \vert V\rangle_b
(|\omega_{1}\rangle|\omega_{2}\rangle
+|\omega_{2}\rangle|\omega_{1}\rangle)_{ab}$,
$\frac{1}{\sqrt{2}}\vert V\rangle_a \vert H\rangle_b
(|\omega_{1}\rangle|\omega_{2}\rangle
+|\omega_{2}\rangle|\omega_{1}\rangle)_{ab}$ or
$\frac{1}{\sqrt{2}}\vert V\rangle_a \vert V\rangle_b
(|\omega_{1}\rangle|\omega_{2}\rangle
+|\omega_{2}\rangle|\omega_{1}\rangle)_{ab}$, respectively. With the
setup shown in Fig. \ref{fig1}, Alice and Bob can get the entangled
state $\vert \psi^+\rangle_{a_1b_1}$, $\vert
\phi^+\rangle_{a_1b_2}$, $\vert \phi^+\rangle_{a_2b_1}$, and $\vert
\psi^+\rangle_{a_2b_2}$ from the output ports $a_1b_1$, $a_1b_2$,
$a_2b_1$, and $a_2b_2$, respectively, as the same as the case with a
collective noise. In a word, this setup for entanglement sharing
works for an arbitrary noise on the polarization states of photons
and the original state in the degree of freedom of polarization can
be an arbitrary one.

\section{Generalization and application of this entanglement distribution scheme over a collective noise}

The present setup can also be used to transmit multi-photon
entangled states faithfully and efficiently. Suppose a multi-photon
system is originally in the state $\vert \Psi\rangle_N =
\frac{1}{\sqrt{2}}\vert H\rangle_1 \cdots \vert H\rangle_{j} \cdots
\vert H\rangle_{N} (\vert \omega_{1} \rangle_1  \cdots \vert
\omega_{1} \rangle_{j} \cdots \vert \omega_{2} \rangle_N  + \vert
\omega_{2} \rangle_1 \cdots \vert \omega_{2} \rangle_{j} \cdots
\vert \omega_{1} \rangle_N )$. Here the subscript $j$
($j=1,2,\cdots, N$) represents the $j$-th qubit which is distributed
to the $j$-th party in quantum communication. Let us assume that the
effect of channel noise on each qubit can be described as $\vert H
\rangle_i \;\; ^{\underline{ noise }}\rightarrow \alpha_j \vert H
\rangle + \beta_j \vert V \rangle$, where $\vert \alpha_j \vert
^{2}+ \vert \beta_j \vert^{2}=1$. After the transmission over the
collective-noise channels, the $N$-qubit system is in the state
\begin{eqnarray}
&&\frac{1}{\sqrt{2}}\{(\alpha_1 \cdots \alpha_{j} \cdots
\alpha_N\vert H\rangle_1  \cdots \vert H\rangle_j \cdots \vert
H\rangle_{N}   +  \cdots \nonumber\\
&& \;\;\;\;\; \; +  \alpha_1 \cdots \beta_{j} \cdots
\alpha_N\vert H\rangle_1  \cdots \vert V\rangle_j \cdots \vert H\rangle_{N} \nonumber\\
&& \;\;\;\;\;\;  +  \cdots \; + \; \beta_1 \cdots \beta_{j} \cdots
\beta_N\vert V\rangle_1  \cdots \vert V\rangle_j \cdots \vert
V\rangle_{N}) \nonumber \\
&& \cdot(  \vert \omega_{1} \rangle_1 \cdots \vert \omega_{1}
\rangle_{j} \cdots \vert \omega_{2} \rangle_N   + \vert \omega_{2}
\rangle_1   \cdots \vert \omega_{2} \rangle_{j} \cdots \vert
\omega_{1} \rangle_N ) \}.\nonumber\\
\end{eqnarray}
With the similar setups shown in Fig.\ref{fig1}, the entangled
photon systems with different polarization states will emit from
different spatial modes and the $N$ parties can obtain an $N$-qubit
entangled state $\frac{1}{\sqrt{2}}( \vert H\rangle_1 \cdots \vert H
\rangle_{j} \cdots \vert H \rangle_N   + \vert V \rangle_1 \cdots
\vert V \rangle_{j} \cdots \vert V \rangle_N )$ by converting the
frequency-entangled state $\frac{1}{\sqrt{2}}(  \vert \omega_{1}
\rangle_1 \cdots \vert \omega_{1} \rangle_{j} \cdots \vert
\omega_{2} \rangle_N   + \vert \omega_{2} \rangle_1   \cdots \vert
\omega_{2} \rangle_{j} \cdots \vert \omega_{1} \rangle_N )$ to a
polarization-entangled one.

Obviously, this setup for faithful entanglement sharing has some
good applications in almost all one-way quantum communication
protocols based on entanglement for rejecting the errors caused by
the collective noise in channels. For example, the BBM92 QKD
protocol \cite{BBM92} can be carried out perfectly through a noisy
channel by using this setup for its entanglement sharing. In this
time, the two parties Alice and Bob first distribute a
polarization-entangled state $\frac{1}{\sqrt{2}}(\vert H\rangle_a
\vert H\rangle_b + \vert V\rangle_a  \vert V\rangle_b)$, resorting
to the frequency-entangled state
$\frac{1}{\sqrt{2}}|H\rangle_{a}|H\rangle_{b}
(|\omega_{1}\rangle|\omega_{2}\rangle
+|\omega_{2}\rangle|\omega_{1}\rangle)_{ab}$, and then they can
choose one of the two nonorthogonal bases, $Z=\{\vert H\rangle,
\vert V\rangle\}$ basis and $X=\{\frac{1}{\sqrt{2}}(\vert H\rangle
\pm \vert V\rangle)\}$ basis, to measure the quantum state of the
two photons emit from the spatial modes $a_1b_1$, $a_1b_2$,
$a_2b_1$, or $a_2b_2$. That is, Alice and Bob measure their photons
from each outport with one of the two bases, as the same as the
original BBM92 QKD protocol. After measurement, they compare their
bases for each entangled photon pair and keep the outcomes when they
choose the correlated bases. Also, this setup for faithful
entanglement sharing  can be used directly in the HBB QSS protocol
\cite{QSS} in which the three parties first share an entangled
three-photon Greenberger-Horne-Zeilinger state
$\frac{1}{\sqrt{2}}(\vert H\rangle_a \vert H\rangle_b \vert
H\rangle_c  + \vert V\rangle_a \vert V\rangle_b \vert V\rangle_c)$
and then each party measures his photon polarization state with one
of the two bases $Z$ and $Y=\{\frac{1}{\sqrt{2}}(\vert H\rangle \pm
i\vert V\rangle)\}$. They can create a private key by the
correlation among the three photons with the suitable bases.

\section{Discussion and summary}

The present setup for faithful entanglement sharing  has several
important characters. First, an arbitrary error on the polarization
state of each qubit caused by the noisy channel can be rejected with
the success probability of 100\% in principle. The significant
hypothesis is that the channel noise is frequency-independent.
Second, it works without resorting to additional qubits
\cite{prl95}, fast polarization modulators \cite{pla343}, and
nondestructive quantum nondemolition detectors (QND) based on
nonlinear media \cite{shengpra}.  Third, its success probability is
independent of the noise parameters and the two parties in quantum
communication can obtain a perfect entangled state by postselection
in a deterministic way, not a probabilistic one \cite{apl91}.
Fourth, this faithful entanglement sharing scheme can be applied
into not only QKD but also QSS based on entanglement.

Certainly, there are some other problems when the present scheme is
used in a practical quantum communication \cite{dengqic}. One is the
stability of the optical paths from the WDM filter to the PBS in
each party of quantum communication, which means that the phase and
the polarization state of the photons in the laboratory of each
party should be held. At present, it is not easy for us to maintain
the stabilization of the optical paths for a long time. On one hand,
this feature will improve the difficulty of the implementation of
the present scheme in a practical application. On the other hand,
the  parties in quantum communication can use some reference signals
to analyze periodically the  stability of the optical paths and
compensate the fluctuation with feedback. With the improvement of
technology, the parties can also use some interferometers with
optical integrations in chips to depress the fluctuation of the
optical paths.  Another problem is the channel losses and detection
dark counts. When the present scheme is applied in QKD or QSS based
on entanglement, it will suffer from the channel losses and
detection dark counts, the same as other faithful qubit transmission
schemes \cite{prl95,xihanOC,pla343,apl91,shengpra,lixhjpb} and
quantum communication protocols \cite{RMP}. In fact, the detection
dark counts will decease the key-generation rate as its effect
equals to lose a portion of the entangled photon pairs transmitted
over a collective-noise channel. This is a common problem in a
practical quantum communication. On the other hand, the channel
losses has two effects. One is that it decreases the key-generation
rate if a photon in an entangled system is lost before it arrives
the side of a party in quantum communication as the parties only
exploit the instances in which all the parties detect a photon and
get a outcome of measurements to exchange their message. The other
is that it will decrease the success probability of the present
entanglement distribution scheme if only some photons in the
entangled quantum system are lost.

The main part of this setup for faithful entanglement sharing is the
entanglement transfer between the frequency degree of freedom and
the polarization degree of freedom. On the one hand, the frequency
difference between the two photons should be set to a small value
which could be discriminated by the WDM efficiently. Also, the small
value of the frequency difference will make the effect of a noisy
channel on  photons with different frequencies be the same one
approximately.  On the other hand, the frequency of the two photons
must be reset to the same value by FS, which is essential to get
standard Bell states. The FS can be implemented by several means
with current technique, such as sum-frequency generation  process
\cite{r12} whose internal conversion efficiency is 99\% \cite{r13}.
At present, the entanglement on the frequency degree of freedom of
photons (called color entanglement) is feasible in experiment
\cite{color}, which makes the present feasible.

In summary, we have presented  a setup for faithful entanglement
sharing against a collective channel noise, without resorting to
additional qubits, fast polarization modulators, and nondestructive
quantum nondemolition detectors based on nonlinear media. This setup
is composed of PBSs,  HWPs, WDMs, and FSs, which is feasible with
current technology. An arbitrary qubit error on the polarization
state of each qubit caused by the noisy channel can be rejected with
the success probability 100\% in principle. Moreover, its success
probability is independent of the noise parameters. This setup has
good applications in  almost all one-way quantum communication
protocols based on entanglement for rejecting the errors caused by
the collective noise in channels. We have discussed its applications
in the famous BBM92 QKD scheme and the HBB QSS protocol.

\section*{ACKNOWLEDGEMENTS}

We would like to thank Dr. Xi-Han Li for helpful discussion. This
work is supported by the National Natural Science Foundation of
China under Grant Nos. 10974020 and 11174039, NCET, and the
Fundamental Research Funds for the Central Universities.

\end{document}